\documentstyle[twocolumn,prl,aps,epsfig]{revtex}
\begin{document}

\title{
Can a frustrated spin-cluster model describe the low-temperature
physics of NaV$_2$O$_5$ ?}

\author{Claudius Gros, Roser Valent\'\i\ and J.V. Alvarez} 
\address{Fachbereich Physik, University of the Saarland,
66041 Saarbr\"ucken, Germany.}
\author{Kay Hamacher and Wolfgang Wenzel} 
\address{   Institut f\"ur
 Physik, Universit\"at Dortmund, 44221 Dortmund, Germany.}
\date{\today}

\maketitle

\begin{abstract}
Recent experimental evidence suggest the existence of three distinct
V-valence states (V$^{+4}$, V$^{+4.5}$ and V$^{+5}$) in the
low-temperature phase of NaV$_2$O$_5$ in apparent discrepancy with
the observed spin-gap. We investigate a novel spin cluster model,
consisting of weakly coupled, frustrated four-spin clusters aligned
along the crystallographic b-axis that was recently proposed to
reconcile these experimental observations. We have studied
the phase diagram and the magnon dispersion relation of this model
using DMRG, exact diagonalization and a novel cluster-operator
theory. We find a spin-gap for all parameter values and two distinct
phases, a cluster phase and a Haldane phase.  We evaluate the size of
the gap and the magnon dispersion and find no parameter regime which
would reproduce the experimental results. We conclude that this model
is inappropriate for the low-temperature regime of NaV$_2$O$_5$.\\

\end{abstract}

{\it Introduction}
Recent investigations of electronically quasi one-dimensional (1D)
transition metal compounds probe the limits of our understanding of
the interplay between structural and electronic effects in such low
dimensional materials. In NaV$_2$O$_5$, a prototypical example for
this class of materials, V-ions are arranged in ladders along the
crystallographic b-direction. Measurements of the magnetic
susceptibility \cite{Isobe96} in the high-temperature phase indicate
the presence of only one equivalent 
V-site \cite{Smolinski98,Schnering98} with valence
V$^{+4.5}$, consistent with a model where
the electrons in bonding V-O-V orbitals along the rungs of the ladder
form a 1D Heisenberg chain\cite{Smolinski98,Horsch98}.

At $T_{C}=34\,\mbox{K}$ the unit-cell doubles along the a- and b- and
quadruples along the c-axis\cite{Fujii97} in a phase transition of
as-of-yet unknown origin. At the same time a spin-gap of
$\Delta_{min}=10\,\mbox{meV}$ opens \cite{Isobe96} and charge ordering
$2\mbox{V}^{+4.5}\rightarrow\mbox{V}^{+4}+\mbox{V}^{+5}$ sets in
\cite{Ohama99}. The observed charge ordering is inconsistent with a generic
spin-Peierls scenario~\cite{Isobe96} and raises the question about the
driving force (lattice, magnetic or Coulomb) responsible for this
transition. Since NaV$_2$O$_5$ is an insulator, the discussion of the
material is simplified by the introduction of pseudospins for the
charge degrees of freedom that couple to the
spin-degrees of freedom~\cite{Seo98,Thalmeier98,Mostovoy98,Deband00}.
The effective spin-Hamiltonian depends, consequently, on the
pattern of charge order \cite{Gros99} and may differ in the 
high- and the low-temperature phase.

\begin{figure}[t]
\epsfig{file=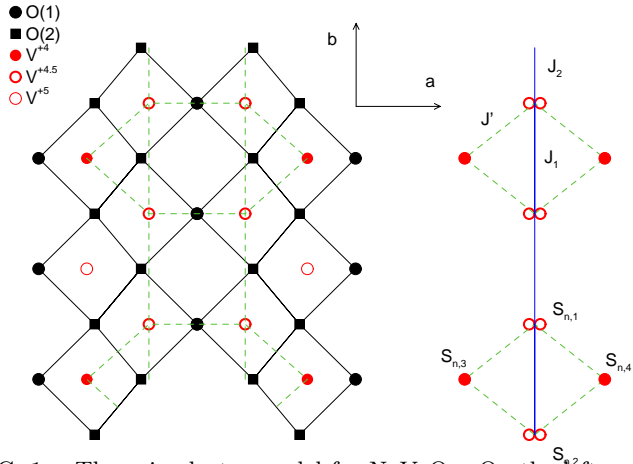,width=0.35\textwidth,angle=-90}
\caption{\label{Fig1}
The spin-cluster model for NaV$_2$O$_5$. On the left
the charge valency in one V-O plane
obtained from two crystal structure determinations
\protect\cite{Smaalen99,Boer99}. On the right the
proposed cluster spin-model \protect\cite{Boer99}.
Note, that two V$^{+4.5}$ on one rung share one
electron.
        }
\end{figure}

The occurrence of two well defined magnon-branches for $T<T_C$ in
NaV$_2$O$_5$ along the $a$ direction (perpendicular to the chains), as
measured by neutron scattering
\cite{Yosihama98}, had been explained tentatively
by a model, where the charge orders in a `zig-zag' pattern in the
low-temperature phase \cite{Gros99}.  This proposal has been
questioned by recent analysis of the low-temperature crystal structure
\cite{Luedecke99,Boer99}. Based on bond-charge models, 
the existence of three different
V-valence states (V$^{+4}$, V$^{+4.5}$ and V$^{+5}$) has been proposed
\cite{Smaalen99,Boer99}, as illustrated in Fig.\ \ref{Fig1}. 
In this analysis,
pairs of V$^{+4.5}$ form dimerized spin-chains on every other
ladder, which alone could explain the  observed spin-gap
\cite{Gros99}. A puzzle is posed however, by the 
presence of free isolated moments on the V$^{+4}$ ions on the
remaining ladders, which is inconsistent with the existence of such a
gap.


As one possible reconciliation, Boer {\it et al.} \cite{Boer99}
recently proposed that the V$^{+4}$ moments are quenched by their
interaction with the neighboring V$^{+4.5}$ sites of the adjacent
dimerized V-O-V ladder.  Within this model, clusters of six Vanadiums
each (and with four spins) would be weakly coupled 
and the observed spin-gap would arise not
from the dimerization but locally from the gap of the isolated
clusters.

To distinguish between these fundamentally different mechanisms we
study this model by a series of complementary approaches, using DMRG
\cite{DMRGBIB}, exact-diagonalization and a novel bond-cluster theory,
to map all physically relevant regions of its phase diagram. We find
that the ground state varies continuously from a cluster-phase for
large $J'$ to a Haldane-phase for small $J'$ (see Fig.\ \ref{Fig1}).
We evaluate the gap and
the dispersion and find that there is no parameter regime that would
explain the neutron-scattering data \cite{Yosihama98}.


{\it The spin-cluster model}
We denote by ${\bf S}_{n,i}$ ($i=1,\dots,4$) the four spins
of the $n$'th cluster, compare Fig.\ \ref{Fig1}. 
The Hamiltonian is then

\[
H= J_1\sum_n {\bf S}_{n,1}\cdot{\bf S}_{n,2}
 +J_2\sum_n {\bf S}_{n,1}\cdot{\bf S}_{n+1,2}
\]
\begin{equation}
+J'\sum_n \left({\bf S}_{n,1}+{\bf S}_{n,2}\right)\cdot
 \left({\bf S}_{n,3}+{\bf S}_{n,4}\right)~,
\label{H}
\end{equation}
where $J_1=J(1+\delta)$ and $J_2=J(1-\delta)$
(with $J_1,J_2,J'>0$). $\delta$
is the degree of dimerization. For $J'=0$ the
${\bf S}_{n,1/2}$ form a dimerized chain
with an in-chain gap $\sim J\delta^{2/3}$. A particular property
of Eq.\ (\ref{H}) is the local coupling to the total
spin ${\bf S}_{n,3}+{\bf S}_{n,4}$, which is consequently
a (locally) conserved quantity,
$({\bf S}_{n,3}+{\bf S}_{n,4})^2=S_n(S_n+1)$ for any $n$. In the
ground-state $S_n\equiv1$. 

We consider first an isolated cluster and denote
by $s_{ij}$ and $t_{ij}^\alpha$ the wavefunctions of
the singlet and of the triplets 
($\alpha=-1,0,+1$) of the spins $i$ and $j$
($i,j=1,\dots4,$).
The low-energy states are 


\begin{equation}
\psi_1 = {1\over\sqrt{3}}\left[
t_{12}^0t_{34}^0 - t_{12}^+t_{34}^- - t_{12}^-t_{34}^+
                         \right]~,
\label{psi1}
\end{equation}
\begin{equation}
\psi_2 = s_{12}^{\phantom{0}}s_{34}^{\phantom{0}},\qquad
\psi_3^\alpha = s_{12}^{\phantom{0}}t_{34}^\alpha~,
\label{psi23}
\end{equation}
\begin{equation}
\psi_4^0={1\over\sqrt{2}}\left[
t_{12}^+t_{34}^- - t_{12}^-t_{34}^+
                         \right]~,
\label{psi4}
\end{equation}
where $\psi_4^0$ is the $S^z=0$ component of the
triplet $\psi_4^\alpha$. The
corresponding energies are
$E_1=-2J'+J_1/4$, $E_2=E_3=-3J_1/4$ and
$E_4=-J'+J_1/4$ \cite{other_states}.

For $J'/J_1>0.5$ the singlet $\psi_1$
is the ground-state (we denote this region
the `cluster phase'). For $J'/J_1<0.5$ the
ground-state of the isolated cluster is
four-fold degenerate, the singlet
$\psi_2$ and the triplet $\psi_3$
have the same energy. Note that the 
intercluster coupling $J_2$ will not
mix $\psi_2$ and $\psi_3$, since the
local spin ${\bf S}_{n,3}+{\bf S}_{n,4}$ is
conserved.
Intercluster coupling $J_2$ will lead to
an antiferromagnetic interaction 
$J_H\sim(J'\,J_2)^2/J_1^3$ between
the moments of the $\psi_3$ states, as can be 
evaluated easily in second-order perturbation in $J_2$ (using
the complete set of eigenstates of the cluster).
The total energy is therefore lowered by $J_2$ when 
all cluster states are $\psi_3$.
The $S=1$ moments of the $\psi_3$ states thus
form an effective spin-1 chain with a Haldane gap
$\Delta_H=0.41050\,J_H$ \cite{White93}.
We denote this region therefore the `Haldane phase'.

\begin{figure}
\epsfig{file=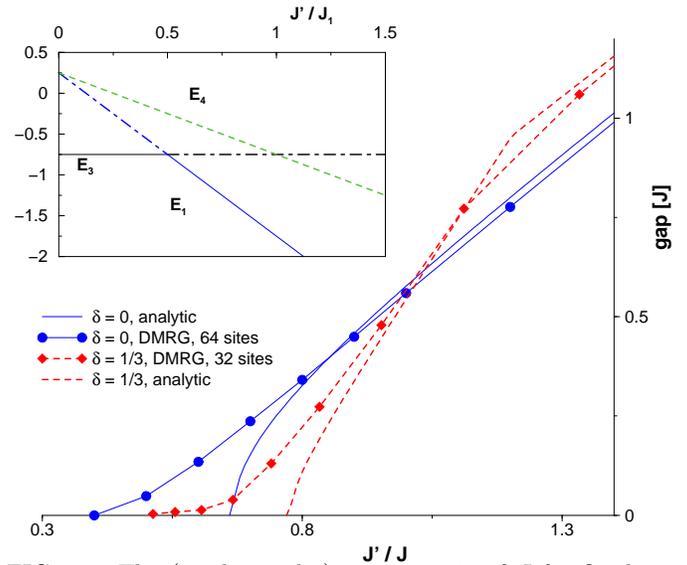,width=0.42\textwidth,angle=-90}

\caption{\label{Fig2}
The (singlet-triplet) gap in units of $J$ for fixed
$\delta=0$ and $\delta=1/3$ as a function of $J'$.
Filled Symbols: DMRG data. Lines: Results from the
cluster operator theory. Inset: The three low-lying
energy levels $E_1$, $E_3$ and $E_4$
for an isolated cluster in units of
$J_1$ as a function of $J'/J_1$. The corresponding
wavefunctions $\psi_3^\alpha$ and $\psi_4^\alpha$ are triplets,
$\psi_1$ is a singlet.
        }
\end{figure}

We have evaluated the energy gap of the spin-cluster
model by DMRG \cite{DMRGBIB}, 
using the finite-size algorithm with
open boundary conditions for systems with $L=32$
and $L=64$ spins. The ground-state has $N_{\uparrow}=L/2$
up-spins and $N_{\downarrow}=L/2$ down-spins. 
We retained typically 60 states
of the density matrix,  checking the convergence
by additional calculations with 40 and 90 states
respectively. We evaluated the gap by two complementary
methods, namely (i) by targeting two states in the
sector with $N_{\uparrow}=L/2=N_{\downarrow}$ and
(ii) by targeting the ground states in 
(a) the sector with $N_\uparrow=L/2=N_\downarrow$
and (b) $N_\uparrow=L/2+1$ and $N_\downarrow=L/2-1$.
We find complete consistency and present the
results in Fig.\ \ref{Fig2} for some selected
values for the dimerization $\delta$. 
The finite-size corrections are smaller than the
symbol sizes. We find a
rapidly decreasing gap as a function of decreasing
$J'/J$ and a smooth crossover between the
cluster- and the Haldane phase. As the symmetry of these
two phases is the same, we do not expect a phase
transition in the thermodynamic limit.

{\it Cluster-operator theory}
In the cluster-phase two low-lying triplet modes,
$\psi_3^\alpha$ and $\psi_4^\alpha$, are relevant. In order to
take the effect of the intercluster coupling $J_2$
into account we describe the seven degrees of
freedom of cluster $n$ by bosonic degrees of freedom:
$s_{n}^\dagger$ for the singlet ($\psi_1$)
$b_{n,3,\alpha}^\dagger$ and
$b_{n,4,\alpha}^\dagger$ for the triplets
($\psi_3$ and $\psi_4$). The low-lying
singlet $\psi_2$ does not couple and may be disregarded
here. This approach generalizes the
bond-operator theory for dimerized spin-chains
\cite{bond_op} to the case of spin-clusters. The
constraint $s_{n}^\dagger s_n^{\phantom{\dagger}}
+ \sum_{\tau,\alpha} 
b_{n,\tau,\alpha}^\dagger b_{n,\tau,\alpha}^{\phantom{\dagger}}
=1$  ($\tau=3,4$) restricts the bosonic Hilbert 
space to the physical one. 
The spin-operators take the form


\begin{equation}
S_{n,1/2}^z = \pm 
 { b_{n,3,0}^\dagger s_{n}^{\phantom{\dagger}}+
        s_{n}^\dagger b_{n,3,0}^{\phantom{\dagger}} 
 \over\sqrt{12}}
-{ b_{n,4,0}^\dagger s_{n}^{\phantom{\dagger}}+
        s_{n}^\dagger b_{n,4,0}^{\phantom{\dagger}} 
 \over\sqrt{6}}~.
\label{S_z}
\end{equation}

Note, that there are no terms 
$\sim b_{n,\tau,\alpha}^\dagger
  b_{n,\tau',\alpha'}^{\phantom{\dagger}}$
corresponding to
triplet-triplet interactions. In 
linearized Holstein-Primakov approximation
(LHP), we substitute $s_{n}^\dagger\rightarrow1$
and $s_{n}^{\phantom{\dagger}}\rightarrow1$
in Eq.\ (\ref{S_z}) and in similar expressions
for $S_{n,1/2}^{x/y}$. This approximation
retains spin-rotational invariance and we may disregard 
the index $\alpha=-1,0,1$ for the triplet operators.
We obtain for the LHP Hamilton-operator in
momentum space $H^{(LHP)}=H_0+H_2^{(1)}+H_2^{(2)}$ with
$H_0 = \sum_{k,\tau}\Delta_\tau
b_{k,\tau}^\dagger b_{k,\tau}^{\phantom{\dagger}}
$
($\Delta_\tau=E_\tau-E_1$). The intercluster
coupling is given by
\begin{figure}
\epsfig{file=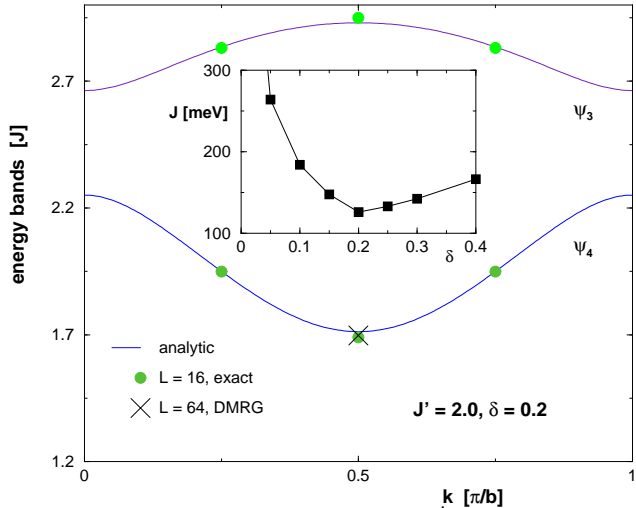,width=0.37\textwidth,angle=-90}

\caption{\label{Fig3}
The magnon dispersion for $J'=2J$,$\delta=0.2$. Note the
zero of energy. The lines are the result of the
cluster-operator theory, the circles of an exact
diagonalization study with 16 spins (periodic boundary
conditions). The cross denotes the DMRG result.
Inset: The value of $J$ as a function of dimerization
$\delta$ needed to fit the measured dispersion of
NaV$_2$O$_5$ (LHP result, for $J/2<J'<J$).
        }
\end{figure}

\[
H_2^{(1)} = {J_2\over12}\sum_{k}\left[ 
2\cos(2bk)\left(
2b_{k,4}^\dagger b_{k,4}^{\phantom{\dagger}}
-b_{k,3}^\dagger b_{k,3}^{\phantom{\dagger}} \right)
\right.
\]
\begin{equation} \left.
+i2\sqrt{2}\,\sin(2bk)
\left(
 b_{k,3}^\dagger b_{k,4}^{\phantom{\dagger}}
-b_{k,4}^\dagger b_{k,3}^{\phantom{\dagger}} \right)
\right]
\label{H_21}
\end{equation} 
and

\[
H_2^{(2)} = {J_2\over12}\sum_{k}\left[ 
\cos(2bk)\left(
2b_{-k,4}^\dagger b_{k,4}^{\dagger}
-b_{-k,3}^\dagger b_{k,3}^{\dagger} \right) \right.
\]
\begin{equation} \left.
-i2\sqrt{2}\,\sin(2bk)\,
b_{k,4}^\dagger b_{-k,3}^{\dagger}\ \ +\ \ \mbox{h.c.}
\right]~.
\label{H_22}
\end{equation} 

\begin{figure}
\epsfig{file=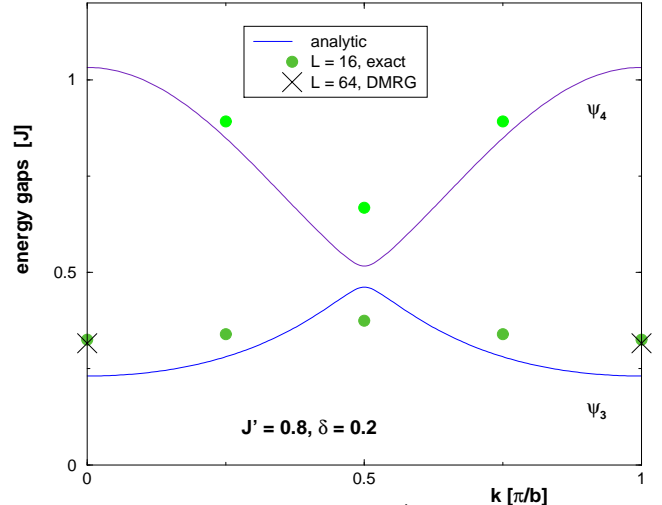,width=0.37\textwidth,angle=-90}
\caption{\label{Fig4}
The magnon dispersion for $J'=0.8\,J$,$\delta=0.2$.The
symbols as in Fig.\ \protect\ref{Fig3}. Note that the
cluster-operator theory overestimates the magnon-dispersion. 
        }
\end{figure}

Here $b=3.611~\mbox{\AA}$ is the lattice constant
of the high-temperature phase.
Note the opposite sign in the dispersion of 
two triplets. It is straightforward
to diagonalize $H^{(LHP)}$.
We define $c=(J_2/6)\cos(2bk)$,
$2t=\Delta_4^2+\Delta_3^2+2c(2\Delta_4-\Delta_3)$
and
$s=\Delta_3^2\Delta_4^2
+2c\Delta_3\Delta_4(2\Delta_3-\Delta_4)
-2\Delta_3\Delta_4 J_2^2/9$.
The dispersion $\omega_\pm=\omega_\pm(k)$
of the two magnon branches (each branch is
three-fold degenerate) in LHP-approximation
is then

\begin{equation}
\omega_\pm^2 = t\pm\sqrt{t^2-s}~.
\label{disp}
\end{equation}
 
We have included the results
for the magnon gap in Fig.\ \ref{Fig2}.
For large ratios $J'/J$ the LHP-result
becomes asymptotically exact, in this limit
it is equivalent to perturbation theory in
$J_2$. In the LHP-approximation the transition to
the Haldane phase is signaled by a vanishing
of the energy gap, the crossover cannot be
described by the cluster-operator theory.

In Fig.\ \ref{Fig3} we present the magnon-dispersion
Eq.\ (\ref{disp}) for $J'=2J$ and compare the LHP-results
(lines) with an exact-diagonalization study of a
system with 16 sites (filled circles) \cite{Muthu96}. 
The agreement is very good, due to the large gap and 
(correspondingly) small correlation length.
Note that the low-lying
magnon, which corresponds to $\psi_4$
(see inset of Fig.\ \ref{Fig2}), 
has its minimum at $k=\pi/(2b)$.

In Fig.\ \ref{Fig4} we present the magnon-dispersion
Eq.\ (\ref{disp}) for $J'=0.8J$ which is closer to the
transition to the Haldane phase. 
The agreement with the exact diagonalization and the
DMRG data is not good, since the precursors
to the Haldane phase are not included in the
cluster-operator theory. The low-lying
magnon, which corresponds to $\psi_3$
(see inset of Fig.\ \ref{Fig2}), has its
minimum now at $k=0$ and $k=\pi/b$ and a maximum 
at $k=\pi/(2b)$, as measured by neutron scattering
\cite{Yosihama98}. The cluster-operator
theory substantially overestimates the size of the
magnon dispersion relative to the exact-diagonalization
result near to the Haldane phase. The physical reason
for this discrepancy can be understood: The lattice
constant of the effective spin-1 chain in the Haldane
phase is $2b$ and the minimum of the magnon dispersion
is therefore at $\pi/(2b)$ 
in the Haldane phase \cite{Affleck89}. It
changes therefore at the crossover from the cluster
phase and the Haldane phase.
This change in the location of the gap is not included
in the cluster-operator theory.


{\it Discussion}
The exchange constant along $b$ is 
$J\approx 529-560\,\mbox{K}$ \cite{Isobe96,Mila96}
in the high-temperature phase of NaV$_2$O$_5$
and the inter-ladder coupling is probably very small,
a $J'/J\approx 1/45$ has been found in an analysis
of the magnon-dispersion for $T<T_{C}$ in a model with
zig-zag charge order \cite{Gros99}. This small
ratio is consistent with the very small coupling along $a$ found
in a LDA-study \cite{Smolinski98}. There are, however,
two reasons why $J'$ might be 
larger in the low-temperature phase. (a) As noted
by Mack and Horsch \cite{Horsch98}, 
there is a near cancellation
for $T>T_C$ in between paths with intermediate
singlet and triplet states and energies $E_{s/t}$:
$J'= t_{xy}^2(1/E_s-1/E_t)$,
where $t_{xy}$ is the V-V hopping matrix element in
$a$ direction. A corresponding calculation for $T<T_C$
in the phase shown in Fig.\ \ref{Fig1} 
yields $J'= 2t_{xy}^2(1/E_s+1/U-1/E_t)$ 
($U$ is the onsite Hubbard-$U$).
(b) 
$t_{xy}$ might be substantially larger in the
low-temperature phase, since the smallness of $t_{xy}$ for
$T>T_C$ is a subtle band-structure effect \cite{Smolinski98}.
We have therefore scanned the complete
phase diagram of the spin-cluster Hamiltonian in order
to determine whether there exists a parameter range
able to fit the neutron scattering data.

We have tried to reproduce, within the
spin-cluster model, four known properties of
NaV$_2$O$_5$: (i) The gap (averaged over $k_a$)
is $\Delta_{min}=10\,\mbox{meV}$. (ii) The maximum
of the dispersion of the lowest magnon branch is at
$\pi/(2b)$, the minimum at $0$ and $\pi/b$.
(iii) The value of the maximum
of the dispersion of the lowest magnon branch
is $\Delta_{max}\approx40\,\mbox{meV}$ 
\cite{Yosihama98,Yosihama99}, i.e.\ the
ratio is $\Delta_{max} / \Delta_{min}\approx4$.
(iv) The value of the coupling along $b$ is
 $J\approx 441\,\mbox{K}=38\,\mbox{meV}$ for $T<T_{SP}$
\cite{Yosihama98,Weiden97}.

Condition (ii) implies that only the cluster-phase
of Hamiltonian Eq.\ (\ref{H}) with $J'<J_1$ is a candidate
for the low-temperature phase of NaV$_2$O$_5$. This
implies  $J_1/2<J'<J_1$.
Within the cluster-operator theory 
one obtains $\Delta_{max} / \Delta_{min}=4$
for values of $J'$ near to the gap closing.
One needs consequently large coupling constants 
$J$ (see inset of Fig.\ \ref{Fig3}).
in order to reproduce $\Delta_{min}=10\,\mbox{meV}$.
We have evaluated the values of $J'$ and $J$ needed
to reproduce the gap-ratio as a function of dimerization
$\delta$ and find a minimum in $J$ for $\delta=0.2$
(see inset of Fig.\ \ref{Fig3}). This minimum is 
$J\approx126\mbox{meV}$, substantially larger than
the experimental value $J\approx 38\,\mbox{meV}$. Note,
that the cluster-operator theory overestimates the
dispersion in this phase and {\bf underestimates}
the value of $J$ needed. We therefore conclude safely,
that the 
model is not able to
reproduce the measured magnon-dispersion of
NaV$_2$O$_5$ and that Eq.\ (\ref{H}) 
is unlikely to be the appropriate model for the
low-temperature phase of NaV$_2$O$_5$, at least in
its one-dimensional version. It might be possible,
in principle, that two-dimensional couplings change
the scenario obtained in the present study, though
we note, that an increase in dimensionality does,
in general, reduce the size of a spin-gap. 


We would like to acknowledge discussions with 
P. Lemmens and the support of the German Science
Foundation, the BMBF and the Fonds der chemischen Industrie.




\begin{thebibliography}{99}

\bibitem{Smolinski98} H. Smolinski {\it et al.},
                      Phys. Rev. Lett. {\bf 80}, 5164 (1998).

\bibitem{Schnering98} H.G. von Schnering {\it et al.},
                      Z. Kristallg. {\bf 213}, 246 (1998);
                      A. Meetsma {\it et al.},
                      Acta Cryst. C {\bf 54}, 1558 (1998).

\bibitem{Horsch98} P. Horsch and F. Mack, 
                   Euro. Phys. J. B {\bf 5}, 367 (1998).

\bibitem{Isobe96} M. Isobe and Y. Udea,
                J. Phys. Soc. Jap. {\bf 65}, 1178 (1996).

\bibitem{Fujii97} Y. Fujii {\it et al.},
                  J. Phys. Soc. Jap. {\bf 66}, 326 (1997).

\bibitem{Ohama99} T. Ohama, H. Yasuoka, M. Isobe and Y. Udea,
                  Phys. Rev. B {\bf 59}, 3299 (1999).

\bibitem{Seo98} H. Seo and H. Fukuyama, 
                J. Phys. Soc. Jpn. {\bf 67}, 2602 (1998).

\bibitem{Thalmeier98} P. Thalmeier and P. Fulde,
                      Europhys. Lett. {\bf 44}, 242 (1998).

\bibitem{Mostovoy98} M. Mostovoy and D. Khomskii, 
                     Sol. St. Comm. {\bf 113}, 159 (1999).

\bibitem{Deband00} Debanand  Sa and C. Gros, preprint.

\bibitem{Gros99} C. Gros and R. Valent\'\i,
                 Phys. Rev. Lett. {\bf 82}, 976 (1999).

\bibitem{Luedecke99} J. L\"udecke, A. Jobst, S. van Smaalen,
                     E. Morr\'e, C. Geibel and H.-G. Krane,
                     Phys. Rev. Lett. {\bf 82}, 3633 (1999).

\bibitem{Boer99} J.L.de Boer, A.M. Meetsma, J. Baas 
                 and T.T.M. Palstra, preprint.

\bibitem{DMRGBIB} S.R. White, Phys. Rev. Lett. {\bf 69}, 2863 (1992). ;
                  I. Peschel {\it et al.} (eds), {\it
                  Density-Matrix Renormalization} (Springer, Berlin, 1999)

\bibitem{Smaalen99} S. van Smaalen and J. L\"udecke,
                     Europhys. Lett. {\bf 49}, 250 (1999).

\bibitem{Yosihama98} T. Yosihama {\it et al.},
                      J. Phys. Soc. Jap. {\bf 67}, 744 (1998).

\bibitem{other_states} The remaining eigenstates are the
                       $S=2$ state $\psi_5^\beta$  
                       ($\beta=-2,\dots,2$) and the triplet 
                        $\psi_6^\alpha=t_{12}^\alpha 
                          s_{34}^{\phantom{\alpha}}$, with energies
                        $E_5=J'+J_1/4$ and $E_6=J_1/4$.

\bibitem{White93} S.R. White and D.A. Huse,
                  Phys. Rev. B {\bf  48}, 3844 (1993).

\bibitem{bond_op}  S. Sachdev and R.N. Bhatt,
                   Phys. Rev. B {\bf 41}, 9323 (199);
                   B. Normand {et al.},
                   Phys. Rev. B {\bf 56}, R5736 (1997);
                   W. Brenig, Phys. Rev. B {\bf 56},
                   14 441 (1997).  

\bibitem{Muthu96} For the method used see 
            V.N. Muthukumar {\it et at.} 
            Phys. Rev. B {\bf 54}, R9635 (1996).

\bibitem{Affleck89} See, e.g., I. Affleck, 
                    J. Phys. Cond. Matt. {\bf 1}, 3047 (1989).

\bibitem{Mila96} F. Mila, P. Millet and J. Bonvoisin,
               Phys. Rev. B {\bf 54}, 11 925 (1996).
 
\bibitem{Yosihama99} T. Yosihama, private communication.

\bibitem{Weiden97} M. Weiden {\it et al.},
                   Z. Phys. B {\bf 103}, 1 (1997).






\end{thebibliography}
\end{document}